\documentclass[prl,nofootinbib,twocolumn,superscriptaddress,preprintnumbers,balancelastpage,longbibliography]{revtex4-1}
\usepackage[colorlinks=true,urlcolor=brown,linkcolor=blue,citecolor=magenta]{hyperref}

\usepackage{graphicx,slashed,xcolor,hepunits,soul}
\usepackage{amssymb}
\usepackage[english]{babel}

\newcommand\bea{\begin{equation}}
\newcommand\eea{\end{equation}}
\def\lsim{\mathrel{\rlap{\lower4pt\hbox{\hskip1pt$\sim$}}
     \raise1pt\hbox{$<$}}}         
\def\gsim{\mathrel{\rlap{\lower4pt\hbox{\hskip1pt$\sim$}}
     \raise1pt\hbox{$>$}}}

\begin{document}
\title{Bounds on Gauge Bosons Coupled to Non-conserved Currents }

\author{Majid Ekhterachian}
\email{ekhtera@umd.edu}
\affiliation{Maryland Center for Fundamental Physics, Department of Physics, University of Maryland, College Park, MD 20742.}

\author{Anson Hook}
\email{hook@umd.edu}
\affiliation{Maryland Center for Fundamental Physics, Department of Physics, University of Maryland, College Park, MD 20742.}

\author{Soubhik Kumar}
\email{soubhik@berkeley.edu}
\affiliation{Berkeley Center for Theoretical Physics, Department of Physics, University of California, Berkeley, CA 94720, USA}
\affiliation{Theoretical Physics Group, Lawrence Berkeley National Laboratory, Berkeley, CA 94720, USA
}
\author{Yuhsin Tsai}
\email{ytsai3@nd.edu}
\affiliation{Department of Physics, University of Notre Dame, IN 46556, USA}

\begin{abstract}
We discuss new bounds on vectors coupled to currents whose non-conservation is due to mass terms, such as $U(1)_{L_\mu - L_\tau}$.
Due to the emission of many final state longitudinally polarized gauge bosons, inclusive rates grow exponentially fast in energy, leading to constraints that are only logarithmically dependent on the symmetry breaking mass term.  
This exponential growth is unique to Stueckelberg theories and reverts back to polynomial growth at energies above the mass of the radial mode.
We present bounds coming from the high transverse mass tail of mono-lepton+missing transverse energy events at the LHC, which beat out cosmological bounds to place the strongest limit on Stueckelberg $U(1)_{L_\mu - L_\tau}$ models for most masses below a keV.
We also discuss a stronger, but much more uncertain, bound coming from the validity of perturbation theory at the LHC.
\end{abstract}

\maketitle
\flushbottom

\section{Introduction}
Recently, new light weakly coupled particles have increasingly become a focus as either a mediator to a dark sector~\cite{Boehm:2003hm,Pospelov:2007mp,ArkaniHamed:2008qn}, as dark matter itself~\cite{Abbott:1982af,Dine:1982ah,Preskill:1982cy,Arias:2012az,Graham:2015rva}, or to explain potential anomalies~\cite{Gninenko:2001hx,Kahn:2007ru,Pospelov:2008zw,TuckerSmith:2010ra,Batell:2011qq,Feng:2016ysn}.  A particularly well motivated candidate is a vector boson.  The currents that these vector bosons couple to can either be conserved, e.g. $U(1)_{B-L}$ with Dirac neutrino masses, or they can be non-conserved, e.g. $U(1)_L$ or $U(1)_{L_\mu - L_\tau}$.

In this letter we will continue a long line of research into the bounds that can be placed on vector bosons coupled to non-conserved currents, see e.g. Refs.~\cite{Preskill:1990fr,Fayet:2006sp,Barger:2011mt,Karshenboim:2014tka,Dror:2017ehi,Dror:2017nsg,Krnjaic:2019rsv,Dror:2020fbh}.  As is well known, these models are non-renormalizable field theories and as a result have amplitudes that grow with energy.  Eventually these amplitudes grow so large that tree level amplitudes violate unitarity at the energy scale $\Lambda$, indicating that perturbation theory has broken down.  Requiring that new physics appears below the scale $\Lambda$ gives the unitarity bound.  Unitarity bounds have a long storied history, see e.g. Refs.~\cite{LlewellynSmith:1973yud,Cornwall:1973tb,Joglekar:1973hh,Cornwall:1974km}, and the famous application to the Higgs boson~\cite{Lee:1977eg,Lee:1977yc,Chanowitz:1985hj,Appelquist:1987cf}.

Typically non-conservation of the currents comes from either anomalies or mass terms.
In the first case, amplitudes involving longitudinal modes are enhanced~\cite{Preskill:1990fr}.  Later,  Ref.~\cite{Dror:2017ehi,Dror:2017nsg} showed that this enhancement led to strong constraints on anomalous gauge theories.
In the second case, inclusive amplitudes were shown to exhibit exponential growth~\cite{Craig:2019zkf}.  We will show that this exponential growth leads to very strong constraints on these theories.

In this letter we will focus on the case of currents that are not conserved due to mass terms.
Specifying this starting point locks us into considering the Stueckelberg limit of gauge theories\footnote{The Stueckelberg limit is when the mass of the radial mode is taken to be heavier than the energy scale under consideration.} as including the radial mode renders mass terms gauge invariant.
A Stueckelberg theory coupled to a non-conserved current can have inclusive rates that grow exponentially fast in energy due to multiparticle emission~\cite{Craig:2019zkf}.  This exponential growth is unique to the Stueckelberg limit and becomes polynomial at energies above the mass of the radial mode.

The exponential growth of amplitudes in these models gives extremely strong unitarity bounds~\cite{Craig:2019zkf}.  Redoing the unitarity bound using our conventions, we find
\bea
\label{eq: unitarity}
\Lambda \approx \frac{4 \pi m_X}{\sqrt{27} g_X} \log^{3/2} \left ( \frac{m_X}{g_X m} \right ),
\eea
where $g_X$ and $m_X$ are respectively the coupling and the mass of the gauge boson, and $m$ is the symmetry breaking mass term.
To show how strong this unitarity bound is, let us consider the Stueckelberg limit of $U(1)_{L_\mu}$.  Typically one assumes that the strongest unitarity bound on this model comes from the fact that this current is anomalous.  Redoing the unitarity bound (for details see the Supplementary Material) using our conventions leads to a slightly stronger result than the standard result~\cite{Preskill:1990fr}
\bea
\Lambda_a \approx \frac{\sqrt{4 \pi} m_X}{g_X} \frac{32 \pi^2}{g^2},
\eea
where $g$ is the $SU(2)_W$ gauge coupling.
Comparing this with Eq.~\eqref{eq: unitarity} taking $m_X/g_X\simeq1$~GeV, motivated by our later results, and $m\simeq0.05$~eV, as the mass of the  neutrino,
we find that $\Lambda \sim \Lambda_a/10$. This shows that despite the extremely small neutrino mass, its non-zero value still gives a unitarity bound almost an order of magnitude more stringent than the anomaly does!  Thus before considering moving to an anomaly free theory such as $U(1)_{L_\mu - L_\tau}$, one should first UV complete the symmetry breaking neutrino mass terms.  In contrast, for a Higgsed $U(1)_{L_\mu}$ one can naturally have the exact opposite scenario where $\Lambda_a \ll \Lambda$.

Unitarity in these models is restored by the inclusion of the radial mode.  However in these UV completions, the small fermion mass is the result of a higher dimensional operator so that scattering of the radial mode has its own unitarity bound.  Constraints on the UV completion are fairly model dependent, however since it is likely that the UV completion only couples to the SM via the neutrino mass term, bounds on it are plausibly fairly weak. On the other hand, for Stueckelberg gauge bosons we obtain robust, model-independent bounds that do not rely on the dynamics of the radial mode. While these bounds can be evaded by going away from the Stueckelberg limit by including the radial mode at sufficiently low energies, the radial mode should then be taken into account whenever processes at energies larger than its mass are studied.

In this letter we will focus on gauge bosons coupled to the lepton number currents, which are only non-conserved because of the extremely small neutrino masses.
To find explicit bounds, we consider the total decay width of the $W$ boson and the high transverse mass tail of mono-lepton+MET (missing transverse energy) events at the Large Hadron Collider (LHC).  
%
The bounds we obtain are essentially independent of what the exact model under consideration is, thus in the introduction we will only list our bounds on $U(1)_{L_\mu - L_\tau}$ with Dirac neutrino masses.  A somewhat uncertain bound of $m_X/g_X \gtrsim 24 \,\, \text{GeV}$ comes from requiring that physics is perturbative at the LHC.
We also find
\bea
\frac{m_X}{g_X} > 1.3 \,\, \text{GeV}; \qquad \frac{m_X}{g_X} > 54 \,\, \text{MeV}
\eea
coming from mono-lepton+MET events at the LHC and the total decay width of the W boson respectively.  In the high mass limit, this is only a factor of $\sim$ 100 weaker than the extremely precise measurements of the $g-2$ of the muon~\cite{Jegerlehner:2009ry,Miller:2012opa}.  Meanwhile for $m_X \lesssim 1$ keV, this is the strongest constraint on these models beating out even cosmological bounds\footnote{If the gauge boson instead coupled to $L_e$, the strongest bounds at low mass are from neutrino oscillations in the earth~\cite{Wise:2018rnb} and it is only for masses $m_X \lesssim 10^{-16}$ eV that our bounds win out.}.

\section{Unitarity bound}


In this section, we derive the unitarity bound on these models, which was also done in Ref.~\cite{Craig:2019zkf} using slightly different techniques\footnote{We use a slightly different definition of the unitarity bound and we consider the process $\nu + n \, \phi \rightarrow \nu + n \, \phi$ rather than $\nu + \overline \nu \rightarrow n \, \phi$, which leads to stronger unitarity bounds.}.
In order to demonstrate the exponential growth of amplitudes, we will consider the toy scenario of a gauge boson coupled to only the left handed piece of a Dirac fermion $\nu$, whose current is not conserved due to explicit breaking by a small Dirac mass term $m_\nu$.  Namely we consider a theory with
\bea\label{eq.scalarnu}
\mathcal{L} = -\frac{1}{4} F_X^2 + i \overline \nu \left ( \slashed{\partial} - i g_X \slashed{A}_X P_L \right ) \nu - m_\nu \overline \nu \nu + \frac{1}{2} m_X^2 A_X^2,
\eea
where $F_X$ is the field strength of the gauge boson $A_X$.
In the limit of small $m_\nu$, the scale at which perturbation theory breaks down, $\Lambda$, is much larger than the mass of the gauge bosons.  In this limit, things simplify as the high energy behavior of $A_X$ can be obtained from the Goldstone boson equivalence theorem. Namely, the matrix element obeys $\mathcal{M}(A_X^L + \cdots) \approx \mathcal{M}(\phi  + \cdots)$, where $A_X^L$ is a longitudinally polarized $A_X$ and $\phi$ is the Goldstone boson.

We obtain the theory with Goldstone bosons by leaving unitarity gauge via a chiral gauge transformation, $\nu_L\rightarrow \exp(i g_X \phi/m_X)\nu_L$.  As the mass term is not gauge invariant, it transforms into
\bea
\label{eq: interactions}
V = m_\nu \overline \nu e^{i g_X P_L \phi/m_X} \nu = \sum_n \overline \nu \frac{m_\nu}{n!} \left( \frac{i g_X P_L \phi}{m_X}\right )^n \nu .
\eea
From this, it is clear that the theory is non-renormalizable and has a UV cutoff.  The Goldstone boson equivalence theorem lets us compute the probability of emitting many longitudinal gauge bosons by calculating the much simpler process of emitting many $\phi$ particles using the above interaction.

As one can see, considering processes involving $n$ $\phi$s gives a matrix element $\sim m_\nu (g_X E/m_X)^n$ that becomes more and more insensitive to the small mass parameter $m_\nu$ as $n$ becomes larger and larger.  This causes the best unitarity bound to come from taking $n$ larger and larger.  However, in the large $n$ limit, the $1/n!$ coming from dealing with identical final state particles penalizes taking $n$ too large.  Thus the optimal unitarity bound comes from taking an intermediate value of the number of gauge bosons $n_{\rm opt}$.

A calculation done in the Supplementary Materials shows that the 
amplitude $\hat{\mathbf{M}}$ of the process $\nu + n \, \phi \rightarrow \nu + n \, \phi$ is
\begin{eqnarray}
\label{Eq: M}
|\hat{\mathbf{M}}(\nu + n \, \phi &\rightarrow& \nu + n \, \phi)| = \nonumber \\ 
&\,& \frac{g_X m_\nu}{2 m_X (n+1)! n! (n-1)!} \left ( \frac{g_X E}{4 \pi m_X} \right )^{2n -1}
\end{eqnarray}
in the limit that $E \gg n \, m_X$.  Unitarity requires that $| \hat{\mathbf{M}}  | < 1$.  
In order to obtain the strongest bounds, we choose the $n$ that maximizes Eq.~\eqref{Eq: M} to obtain in the large $n$ limit
\begin{eqnarray}
\label{Eq: M2}
|\hat{\mathbf{M}}(\nu + n_{\rm opt} \phi \rightarrow \nu + n_{\rm opt} \phi) | \sim \frac{g_X m_\nu}{2 m_X} e^{3 \left ( \frac{g_X E}{4 \pi m_X} \right )^{2/3}}.
\end{eqnarray}
This maximum value of $|\hat{\mathbf{M}}|$ is obtained for $n_{\rm opt} \approx (g_X E/4 \pi m_X)^{2/3}$.
Requiring unitarity holds for Eq.~\eqref{Eq: M2} gives the leading logarithmic behavior
\bea
\label{eq: un_example}
E = \Lambda \approx \frac{4 \pi m_X}{\sqrt{27} g_X} \log^{3/2} \left ( \frac{m_X}{g_X m_\nu} \right ).
\eea

From this calculation we see the behavior claimed in the introduction.
Amplitudes have an exponential growth in energy and the strongest growth comes from emitting multiple gauge bosons.
In this calculation, we made the approximation that the energy carried by each of the $A_X$ gauge bosons is much larger than the mass when we utilized the Goldstone boson equivalence theorem.  Combining the expression for $n_{\rm opt}$ with Eq.~\eqref{eq: un_example} and the requirement that $E \gg n_{\rm opt} \, m_X$, we find that our massless approximation of the unitarity bound is valid when $g_X \lesssim 4 \pi \sqrt{\log ( m_X/g_X m_\nu )/3}$.  In the large log limit, the massless limit is always a valid approximation.

\section{Models}

We now briefly describe the models under consideration and set up some notation.  The results in the next section will be given in the 1-flavor approximation so we will also discuss how to easily take into account the standard 3-flavor set up.  For ease of expression, in this section we will use Weyl notation for fermions.

As mentioned before, we will be considering the Stueckelberg limit of different $U(1)$ gauge theories. We first consider $U(1)_{L_\mu-L_\tau}$.
We assume Dirac neutrinos and that the right-handed neutrinos are neutral under $L_\mu-L_\tau$. 
The flavor basis is related to the mass basis by $\nu_F = U \nu_M$, where $\nu_F$ ($\nu_M$) are the flavor (mass) basis left handed neutrinos and $U$ is the PMNS matrix.  In the flavor basis the neutrino mass term is $\nu^c M_d U^\dagger \nu_F$ where $M_d$ is a diagonal matrix of the neutrino masses $m_{1,2,3}$.  To leave unitary gauge, the flavor-basis SM neutrinos are rotated by,
\begin{align}
    \nu_F \rightarrow P \nu_{F} \quad P = \text{diag}\left(1,e^{+i g_X \phi/m_X},e^{-i g_X \phi/m_X}\right) .
\end{align}
Thus the neutrino mass term involving the Goldstone bosons $\phi$ becomes
\begin{eqnarray}
\mathcal{L}^D_{\nu \, \rm mass} &=& \nu^c M_d U^\dagger P \nu_F + \rm{h.c.} \nonumber \\
&\supset& \sum_{n,j} \frac{1}{n!} \left ( \frac{i g_X \phi}{m_X} \right)^n \nu^c_j M_{d,j} \left ( U^\dagger_{j \mu} \nu_\mu + (-1)^n U^\dagger_{j \tau} \nu_\tau \right ). \nonumber
\end{eqnarray}
From this, we see that any 1-flavor process involving $n$ high energy gauge bosons can be converted into the 3-flavor result by replacing
\begin{align}
\label{Eq: sub1}
m_\nu^2 \rightarrow \sum_{j=1}^{3} \left(|U_{\mu j}|^2 + |U_{\tau j}|^2\right) m_j^2. 
\end{align}

The other gauge theory we will consider is $U(1)_L$ with Majorana neutrino masses.  In this case, after leaving unitary gauge the mass term is
\begin{eqnarray}
\mathcal{L}^M_{\nu \, \rm mass} &=& e^{2 i g_X \phi/m_X} \nu^T_M M_d \nu_M + \rm{h.c.}
\end{eqnarray}
From this, the 1-flavor results can be generalized using the substitution
\begin{align}
\label{Eq: sub2}
g_X \rightarrow 2 \, g_X, \qquad m_\nu^2 \rightarrow  \sum_{l=e,\mu,\tau} \, \sum_{j=1}^{3} |U_{l j}|^2 m_j^2.
\end{align}

\section{Constraints}
\label{Sec:constraints}

\begin{figure}[t]
    \centering
    \includegraphics[width=8cm]{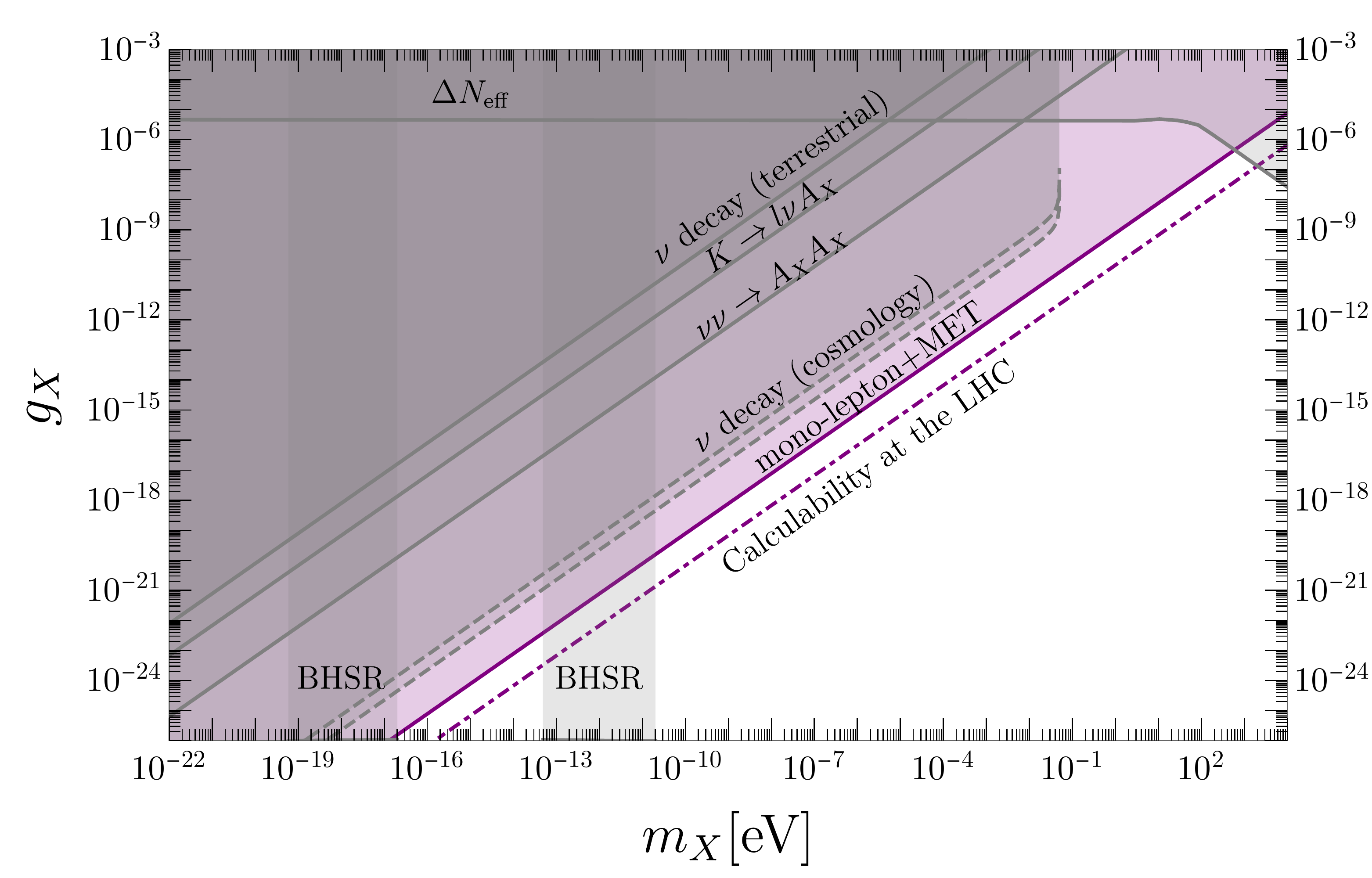}
    \caption{Low mass constraints on a Stueckelberg $U(1)_{L_\mu-L_\tau}$ gauge boson with Dirac neutrino masses. The purple region shows the constraint derived in Eq.~\eqref{eq:newlhcconstraint} coming from the high transverse mass tail of mono-lepton+MET events at the LHC.  The dot-dashed purple line denotes the constraint obtained from demanding the validity of perturbation theory at the LHC, given by Eq.~\eqref{eq:unitarityconstraint}. Other constraints are from: $\Delta N_{\rm eff}$ during BBN through thermalized $A_X$~\cite{Huang:2017egl}, black hole superradiance (BHSR) instability~\cite{Baryakhtar:2017ngi}, rare $K$ decays~\cite{Dror:2020fbh}, $\Delta N_{\rm eff}$ through $\nu\nu\rightarrow A_X A_X$~\cite{Dror:2020fbh,Huang:2017egl}, constraints on $\nu$ decay through terrestrial experiments~\cite{Beacom:2002cb,Funcke:2019grs,Dror:2020fbh}. We also update the cosmological constraints on $\nu$ decay as discussed in the main text, based on the recent result of Ref.~\cite{Barenboim:2020vrr}. Given the $\nu$ decay bounds from cosmology are model-dependent and can be relaxed, we show it via dashed lines.}
    \label{fig:mutau}
\end{figure}
\begin{figure}[t]
    \centering
    \includegraphics[width=8cm]{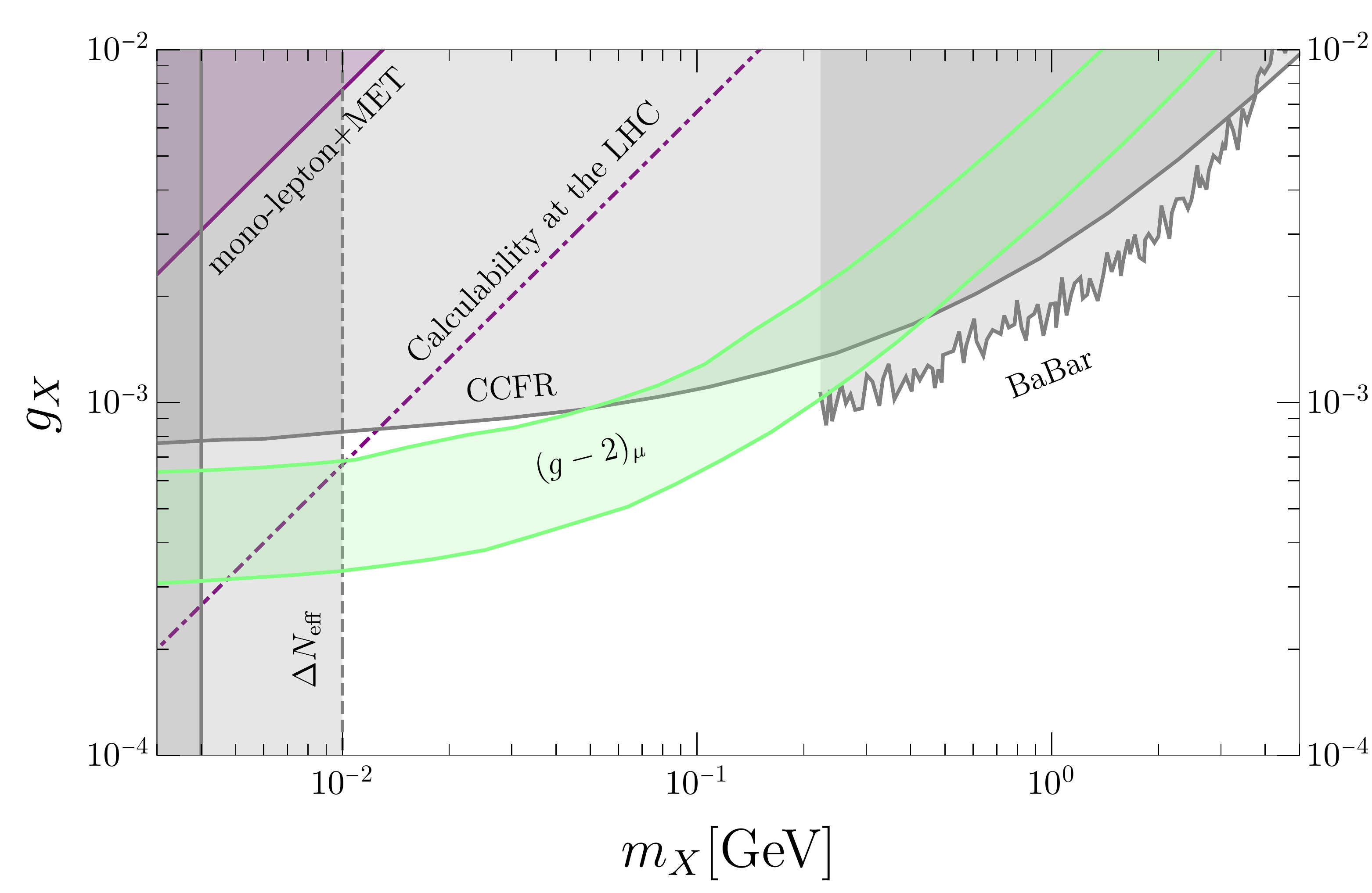}
    \caption{High mass constraints on a Stueckelberg $U(1)_{L_\mu-L_\tau}$ gauge boson with Dirac neutrino masses. The solid and dot-dashed purple lines are the same as in Fig.~\ref{fig:mutau}. 
    Other constraints are from: $\Delta N_{\rm eff}$ during BBN due to thermalization of $A_X$ and entropy dump during its decay~\cite{Huang:2017egl,Kamada:2015era,Escudero:2019gzq} (solid: $T_{\rm RH}>4$~MeV, see e.g.~\cite{Hannestad:2004px,deSalas:2015glj}; dashed: $T_{\rm RH}>10$~MeV), neutrino trident process at the CCFR experiment~\cite{PhysRevLett.66.3117,Altmannshofer:2014pba}, and a search for $e^+ e^-\rightarrow\mu^+\mu^-A_X,A_X\rightarrow\mu^+\mu^-$ at BaBar~\cite{TheBABAR:2016rlg}. In the shaded green region, the muon $(g-2)$ anomaly~\cite{Bennett:2006fi,Aoyama:2020ynm} can be explained by $A_X$ at 2$\sigma$~\cite{Baek:2001kca,Gninenko:2001hx}.}
    \label{fig:mutau2}
\end{figure}

We now present exclusions coming from the emission of many final state gauge bosons.  We will consider three different constraints coming from the total decay width of the $W$ boson, the tail of high transverse mass mono-lepton+MET events, and a rough estimate of when perturbativity breaks down at the LHC.
Our results will be presented in the simplified 1-flavor model. Eq.~\eqref{Eq: sub1} and Eq.~\eqref{Eq: sub2} can be used to convert the results into the two specific models of interest. In particular, in Fig.~\ref{fig:mutau} and Fig.~\ref{fig:mutau2} we show the results after incorporating the full 3-flavor set-up.

\paragraph{W Boson decay : }
We will consider the partial decay width of the W boson into leptons and $n$ gauge bosons
\bea
\Gamma_n (W^- \rightarrow L^- + \overline \nu + n \, A_X).
\eea
The constraint we will impose is that the sum of these decay widths is less than the total decay width
\bea
\Gamma_\text{BSM}\equiv\sum_{n=2}^\infty \Gamma_n < \Gamma_W .
\eea
For simplicity we will take $n \ge 2$ to avoid the soft divergence.

Using the Goldstone boson equivalence theorem, we find the result to leading order in $m_\nu$ to be
\bea
\Gamma_\text{BSM} = \frac{g_2^2 m_\nu^2}{1536 \pi M_W} x^4 \,_2 F_4 ( \{ 1,1 \}, \{ 2,3,3,5 \}, x^2)
\eea
where $x = \frac{g_X M_W}{4 \pi m_X}$ and $\,_p F_q$ is the generalized hypergeometric function.  In the limit of large argument, it simplifies as
\bea
\qquad \,_2 F_4 ( \{ 1,1 \}, \{ 2,3,3,5 \}, x^2) \approx \frac{16 \sqrt{3}}{\pi x^{20/3}} e^{3 x^{2/3}}
\eea
showing the exponential growth of the amplitudes mentioned earlier.

Bounds are obtained by requiring that this decay width is smaller than the total decay width of the W boson.  For $U(1)_{L_\mu-L_\tau}$ ($U(1)_L$) we find that $m_X/g_X > 54$ MeV  ($m_X/g_X > 108$ MeV).  
Because of the exponential, our results are insensitive to the details with which we constrain the model.  For example, requiring that this decay width is smaller than $10^{-6}$ of the total decay width of the W boson (roughly 1 over the number of W bosons produced at LEP) tightens the above $U(1)_{L_\mu-L_\tau}$ bound to $76$ MeV.

\paragraph{mono-lepton+MET events :}


In the next two sub-sections, we will discuss bounds that stem from the breaking of perturbativity at the LHC.
From the decay of the $W$ boson, we see that we are faced with a theory which is becoming dominated by a large number of final state particles.
For black holes~\cite{Banks:1999gd}, when the cross section becomes dominated by a large number of final states, the cross section for s-channel two-to-two scattering becomes highly suppressed.  
The non-observation of this suppression leads to a constraint.    As illustrated below, it is plausible that a similar feature exists in our case when perturbation theory breaks down

The case that we will consider is $p p \rightarrow W^\star \rightarrow l \nu$ via an off-shell $W$ boson.  At high enough energies, the tree level off-shell width of the $W$ boson becomes larger than the momentum flowing through the propagator, suppressing the high transverse mass tail of mono-lepton+MET events~\footnote{Perturbation theory has broken down when this occurs, so that our calculation is only really an estimate of what happens in the full non-perturbative theory.}.  We obtain a bound by requiring that this suppression is small enough that the suppression was not observed in Ref.~\cite{CMS-PAS-EXO-19-017}.

At high energies, the propagator of the $W$ boson is given by
\bea
\frac{i}{s - M_W^2 + \Sigma(s)} \qquad \text{Im}(\Sigma(s)) = M_W \Gamma_W(s),
\eea
where the earlier calculation of the decay width of the $W$ boson gives
\bea
M_W \Gamma_W(s) = \frac{g_2^2 m_\nu^2}{1536 \pi} x^4 \,_2 F_4 ( \{ 1,1 \}, \{ 2,3,3,5 \}, x^2)
\eea
with $x = \frac{g_X \sqrt{s}}{4 \pi m_X}$.  When $ s \ll M_W \Gamma_W(s)$, mono-lepton+MET production is highly suppressed.

Ref.~\cite{CMS-PAS-EXO-19-017} observes the high $M_T$ tail of the mono-lepton+MET events~\footnote{The transverse mass $M_T$ is defined as $M_T=\sqrt{2p_{T}^l p_{T}^{\rm miss}(1-\cos\Delta\phi)}$, where $\Delta\phi$ is the azimuthal angle between $\vec{p}_{T}^l$ and $\vec{p}_{T}^{\rm miss}$.}
to be the expected SM result when $M_T\lesssim 2$~TeV. Given $M_T\leq \sqrt{s}$ by definition, it implies that at center of mass energies of at least  2~TeV, we do not see any deviation from the SM expectation. The observed events dominantly come from $p p \rightarrow W^\star \rightarrow l \nu$.  The off-shell $W$ contribution would be suppressed by at least a factor of 2 if $\sqrt{2} s = M_W \Gamma_W(s)$ and would have been observed for any $M_T \lesssim 2$ TeV.  We obtain a bound by requiring that $\sqrt{2} s \gtrsim M_W \Gamma_W(s)$ at 2 TeV.  For $U(1)_{L_\mu-L_\tau}$  we find that
\begin{align}\label{eq:newlhcconstraint}
m_X/g_X > 1.3~\text{GeV}.    
\end{align}
For $U(1)_L$ with Majorana masses we find $m_X/g_X > 2.6$~GeV.

\paragraph{Calculability : }

The next bound we place is that perturbation theory at the LHC is valid.  This is necessarily a somewhat fuzzy bound because the breakdown of perturbation theory is not very well defined.  For example, the unitarity bound derived in Ref.~\cite{Craig:2019zkf} is weaker than ours by a factor of $\sim 3$ indicating that different techniques give $\mathcal{O}(1)$ different estimates of when perturbation theory breaks down.  However, it is undeniable that we can calculate observables at the LHC and so perturbation theory must be valid.

We will use the unitarity bounds given in Eq.~\eqref{eq: un_example} as an order of magnitude estimate for when perturbation theory breaks down.  The highest center of mass energy collision at the LHC was one that had a center of mass energy of $\sim 8$ TeV~\cite{Sirunyan:2019vgj}.  Requiring that LHC processes are calculable at a center of mass energy of 8 TeV gives the constraint \begin{align}\label{eq:unitarityconstraint}
m_X/g_X \gtrsim 24 ~\text{GeV} 
\end{align}
for $U(1)_{L_\mu-L_\tau}$.
As mentioned before, the breakdown of perturbation theory is a somewhat nebulous concept and thus this bound should be considered as only an estimate of where the bound coming from the breakdown of perturbation theory must lie.

An astute reader will notice that the previous bounds are actually slightly weaker than the unitarity bound found using their relevant energy scales.  As our previous bounds came from tree level calculations, they requires that perturbation theory is valid.  As such, strictly speaking, they only apply if the stronger calculability constraint has misestimated the breakdown of perturbation theory by a factor of few.

We show our constraints visually in Fig.~\ref{fig:mutau} and Fig.~\ref{fig:mutau2} for $U(1)_{L_\mu-L_\tau}$.   We take normal ordering of the neutrino masses with the lightest neutrino being massless.
As explained in the text, the bounds are not sensitive to the precise value of the mass. For other neutrino mixing parameters, we use the results from Table~3 of Ref.~\cite{Esteban:2020cvm}.

In Fig.~\ref{fig:mutau} we have updated the cosmological bounds on these models coming from the decay of the heavier neutrinos, which requires $\tau_0\gsim 4\times 10^{5-6}~{\rm sec}(m_{3}/50~{\rm meV})^5$ ~\cite{Barenboim:2020vrr}. The two $\nu$ decay (cosmology) bounds represent the uncertainty in the bound on the $\nu$ lifetime.  After refining the calculation of the damping of  anisotropic stress due to neutrino decay and inverse decay, this new bound was obtained and found to be several orders of magnitude weaker than previous work (e.g.,~\cite{Archidiacono:2013dua,Escudero:2019gfk}).  For the $\Delta N_{\rm eff}$ constraints, we take $T_{\rm RH} < m_\mu$.  Additionally, in both figures we have neglected constraints coming from a possible coupling to electrons as those bounds are model dependent and can be avoided. 


\section{UV completion}

Unitarity in the emission of gauge bosons can be  restored by introducing the radial mode.  To see the behavior of a UV completion, we consider a complex scalar $\Phi$ with charge 1 and give the neutrinos a charge $q = g_X/g$ so that the coupling of $A_X$ to neutrinos is given by $g_X$ while the coupling of $A_X$ to $\Phi$ is given by $g$.  The symmetry breaking mass term comes from the higher dimensional operator
\bea
\label{eq: mnu}
V = \frac{y \Phi^q}{\Lambda'^{q}} H L \nu^c \qquad m_\nu = \frac{y \, v f^q}{\Lambda'^{q}},
\eea
which gives the standard neutrino mass term after $\Phi$ obtains a vev $f$.
The radial mode can be partially decoupled by taking $g \rightarrow 0$ and $q,f \rightarrow \infty$ while holding $m_X = g f$ and $g_X = q g$ constant.
This limit attempts to decouple the radial mode $m_\Phi \lesssim f$ but at the same time sends the symmetry breaking mass term to zero, $m_\nu \rightarrow 0$.   Thus, when the neutrino mass is non-zero, the radial mode cannot be decoupled.

Using Eq.~\eqref{eq: mnu}, one can show that the unitarity bound $\Lambda$ satisfies
\bea
\Lambda \approx \frac{4 \pi m_X}{\sqrt{27} g_X} \log^{3/2} \left ( \frac{m_X}{g_X m_\nu} \right ) > m_\Phi
\eea
showing that unitarity did indeed predict the correct scale of new physics in these models.  In this UV theory, one can easily show that scattering involving gauge bosons no longer grows.  However, the price is that scattering involving $\Phi$ does grow.  Thus this UV completion will itself require a UV completion at the scale $\Lambda'$.  As this particular higher dimensional operator is identical to those seen in Froggatt–Nielsen models~\cite{Froggatt:1978nt}, its UV completion proceeds along a manner completely analogous to those models and will not be expounded upon here.

It is worth mentioning that while the UV completion presented above can easily obtain $m_\Phi \gg 4 \pi m_X/g_X$, it cannot saturate $\Lambda \sim m_\Phi$.  In this model, $m_\Phi \sim \sqrt{\lambda} f$ which can satisfy $m_\Phi > m_X/g_X \sim f/q$ while perturbation theory is still under control $\lambda q \lesssim 1$.  However in this limit $\Lambda \sim \sqrt{q} f > m_\Phi$, so that saturating the unitarity bound is not possible.  Even when one allows $\lambda q > 1$ and instead uses semi-classical methods~\cite{Badel:2019oxl}, one still finds that it is impossible to saturate the unitarity bound in this model.

The importance of multiparticle emission in Stueckelberg theories could have been anticipated from this UV completion.  In the UV completion, the symmetry breaking mass term arises from a higher dimensional operator of dimension $\sim q$, see Eq.~\eqref{eq: mnu}.  Discovering the bad high energy behavior of a higher dimensional operator of dimension $\sim q$ requires $\sim q$ states.  The IR Stueckelberg theory should match the UV Higgs theory at the scale of the radial mode.  
However, the IR Stueckelberg theory does not know which UV Higgs theory to match onto.  Thus all it can do is at a given energy scale $E$, match onto the UV Higgs theory which has $m_\Phi = E + \epsilon$.  As the energy scale $E$ increases, the IR theory has to match onto different UV theories with larger and larger $m_\Phi$ and hence larger and larger $q$.  Thus when scattering particles at higher and higher energies, the dominant final state involves an ever increasing number of particles.

\section{Conclusion}

In this letter, we considered Stueckelberg gauge bosons coupled to non-conserved currents broken by mass terms and calculated the bounds on these models coming from mono-lepton+MET events at the LHC.  For large gauge boson masses, these constraints are only a bit weaker than even the strongest of bounds, such as $g-2$ experiments.  For most masses below a keV, these constraints are the strongest bounds on these models.

The strength of these bounds comes from the exponential growth of inclusive rates, a feature only present in the Stueckelberg limit~\cite{Craig:2019zkf}.  This growth is a double edged sword.  On one hand, it allows one to use very crude measurements to place extremely stringent constraints.  On the other hand, it does not benefit from precision measurements so that it is not easy to improve on the constraints without access to a higher energy environment.
For example, a 100 TeV collider would improve upon the LHC bounds by about an order of magnitude.

\paragraph{\bf{Acknowledgments}}
We thank Simon Knapen and 
Maxim Pospelov for comments on the draft and Zhen Liu for useful discussions.
ME and AH were supported in part by the NSF grants PHY-1914480, PHY-1914731, and by the Maryland Center for Fundamental Physics (MCFP).
SK was supported in part by the NSF grant PHY-1915314 and the U.S. DOE Contract DE-AC02-05CH11231.
YT was also supported in part by the NSF grant PHY-2014165.

\bibliography{unitarity}

\clearpage
\newpage
\maketitle
\onecolumngrid
\begin{center}
\textbf{\large Bounds on Gauge Bosons Coupled \\
to Non-conserved Currents } \\ 
\vspace{0.05in}
{ \it \large Supplementary Material}\\ 
\vspace{0.05in}
{}
{Majid Ekhterachian, Anson Hook, Soubhik Kumar, Yuhsin Tsai}

\end{center}
\setcounter{equation}{0}
\setcounter{figure}{0}
\setcounter{table}{0}
\setcounter{section}{1}
\renewcommand{\theequation}{S\arabic{equation}}
\renewcommand{\thefigure}{S\arabic{figure}}
\renewcommand{\thetable}{S\arabic{table}}
\newcommand\ptwiddle[1]{\mathord{\mathop{#1}\limits^{\scriptscriptstyle(\sim)}}}

This Supplementary Material contains additional calculations supporting the results in the main text.

\section{Unitarity bounds - $n$-to-$n$ scattering}

When discussing our unitarity bounds, we will follow the conventions and discussions of Ref.~\cite{Chang:2019vez,Abu-Ajamieh:2020yqi} and for simplicity we will be working in the limit of massless particles.
Any S matrix can be decomposed into $S = \mathbf{1} + i T$.  The identity matrix describes the situation where particles pass by without interacting  while the transition matrix $T$ describes nontrivial processes.  
The states we will be considering have a continuous label $P$, the total momentum, and discrete labels $\alpha$.  These states are normalized as
\bea
\label{Eq: norm}
\langle P', \alpha' \mid P, \alpha \rangle = ( 2 \pi)^4 \delta^4 ( P - P' ) \delta_{\alpha \alpha'} .
\eea
The amplitude $\hat{\mathbf{M}}$ is defined by
\bea
\langle P', \alpha' \mid T \mid P, \alpha \rangle = ( 2 \pi)^4 \delta^4 ( P - P' ) \hat{\mathbf{M}}_{\alpha \alpha'} \qquad \qquad \langle P', \alpha' \mid S \mid P, \alpha \rangle = ( 2 \pi)^4 \delta^4 ( P - P' ) S_{\alpha \alpha'} .
\eea
The main result that we will use is that $| \hat{\mathbf{M}}_{\alpha \alpha'} | \le 1$ for all $\alpha$ and $\alpha'$ at tree level.  When $\alpha \ne \alpha'$, this statement follows directly from the conservation of probability.  For $\alpha = \alpha'$ we use unitarity
\bea
1 = \delta_{\alpha \alpha} = \sum_\gamma S^\dagger_{\alpha \gamma} S_{\gamma \alpha} = 1 - 2 \, \text{Im} \, \hat{\mathbf{M}}_{\alpha \alpha} + \sum_\gamma | \hat{\mathbf{M}}_{\gamma \alpha}|^2 .
\eea
From this we have
\bea
2 \, \text{Im} \, \hat{\mathbf{M}}_{\alpha \alpha} = \sum_\gamma | \hat{\mathbf{M}}_{\gamma \alpha}|^2 \ge | \hat{\mathbf{M}}_{\alpha \alpha} |^2 
\eea
which can be massaged into the form $1 \ge | \text{Re} \, \hat{\mathbf{M}}_{\alpha \alpha} |^2 + | \text{Im} \, \hat{\mathbf{M}}_{\alpha \alpha} -1 |^2$.  From this we see that 
$| \text{Re} \, \hat{\mathbf{M}}_{\alpha \alpha} | \le 1$.  Since $\hat{\mathbf{M}}_{\alpha \alpha}$ is real at tree level, we have 
$|\hat{\mathbf{M}}_{\alpha \alpha} | \le 1$.

We now have the unitarity bound $| \hat{\mathbf{M}}_{\alpha \alpha'} | \le 1$ so we can apply it to the theory described in Eq.~\eqref{eq.scalarnu}.  We will be considering the initial and final states each with a neutrino $\nu$ and $n$ Goldstone bosons $\phi$.  We define our states as
\bea
\mid P , n , \alpha \rangle = C_n \int d^4x e^{- i P x} \phi^{(-)}(x)^n \nu_\alpha^{(-)} (x) \mid 0 \rangle 
\eea
where $C_n$ is a normalization constant, $(-)$ are the part of the fields that contain the creation operators, and $\alpha$ is the spinor index of the fermion.  These states are normalized as 
\bea
\langle P', n', \dot \alpha \mid P, n , \alpha \rangle = ( 2 \pi)^4 \delta^4 ( P - P' ) \delta_{n n'} \frac{\slashed{P}^{\alpha \dot \alpha}}{E} 
\eea
where $E$ is the total energy.  This normalization was chosen to reproduce Eq.~\eqref{Eq: norm} in the center of mass frame, which we will be using from here on.  From this, we have the normalization constant
\bea
\frac{1}{|C_n|^2} = \frac{1}{2 (n+1) (n-1)!} \left ( \frac{E}{4 \pi} \right )^{2n-1}.
\eea
We can finally calculate the amplitude of interest using Eq.~\eqref{eq: interactions} to give
\begin{eqnarray}
\langle P', n, \alpha \mid (- i) \int d^4 x \overline \nu(x) \frac{m_\nu}{(2n)!} \left( \frac{i g_X P_L \phi(x)}{m_X}\right )^{2n} \nu(x) \mid P, n , \alpha \rangle &=& ( 2 \pi)^4 \delta^4 ( P - P' ) i \hat{\mathbf{M}}(\nu + n \phi \rightarrow \nu + n \phi) \\
|\hat{\mathbf{M}}(\nu + n \phi \rightarrow \nu + n \phi)| &=& \frac{g_X m_\nu}{2 m_X (n+1)! n! (n-1)!} \left ( \frac{g_X E}{4 \pi m_X} \right )^{2n -1}
\end{eqnarray}
Imposing $|\hat{\mathbf{M}}| \le 1$ gives the results shown in the text.
For reference, the amplitude $\hat{\mathbf{M}}$ is related to the more familiar matrix element $\mathcal{M}$ by normalization constants and phase space integrals
\begin{eqnarray}
\hat{\mathbf{M}}_{\alpha \alpha'} = C^\star_\alpha C_{\alpha'} \int d\Phi_\alpha d\Phi_{\alpha'} \mathcal{M}_{\alpha \alpha'}.
\end{eqnarray}
The phase space differentials $d\Phi_\alpha$ will be given below.

\section{Unitarity bounds on an anomalous $U(1)$}

In this section we calculate the unitarity bounds on an anomalous gauge theory in a manner analogous to what we did for $n$-to-$n$ scattering.  We leave unitarity gauge by a gauge transformation $\phi/m_X$.  Because the theory is anomalous, an anomaly term is added to the Lagrangian
\bea
\mathcal{L} \supset \frac{g^2 \mathcal{A}}{32 \pi^2} \frac{g_X \phi}{m_X} W^a \tilde W^a,
\eea
where $\tilde W^{\mu \nu} = \frac{1}{2} \epsilon^{\mu \nu \rho \sigma} W_{\rho \sigma}$ and $W^a$ are the gauge bosons with which $U(1)_X$ is anomalous and $\mathcal{A}$ is the anomaly coefficient.

We will consider 2-to-2 scattering of $W^1$ gauge bosons via the Goldstone boson $\phi$, which contains all of the leading high energy behavior in Feynman-'t Hooft gauge.  
The largest amplitude occurs when all four gauge bosons have the same  helicity.
A short calculation gives the amplitude
\begin{eqnarray}
|\hat{\mathbf{M}}(W^1 + W^1 \rightarrow W^1 + W^1)| &=& \frac{s}{4 \pi} \left(  \frac{g_X}{m_X} \frac{g^2 \mathcal{A}}{32 \pi^2} \right ) ^2
\end{eqnarray}
Requiring the unitarity bound $|\hat{\mathbf{M}}| < 1$ be saturated at the center of mass energy $\Lambda_a$, we arrive at the result
\bea
\Lambda_a = \frac{\sqrt{4 \pi} m_X}{g_X} \frac{32 \pi^2}{g^2 \mathcal{A}}.
\eea

\section{Decay width of the $W$ boson}
Here we compute the decay width of the $W$ boson into a lepton, a neutrino and $n$ gauge bosons.
The decay is dominated by the decay into longitudinal modes which we calculate using the Goldstone boson equivalence theorem.

\begin{figure}[h]
    \centering
    \includegraphics[width=8cm]{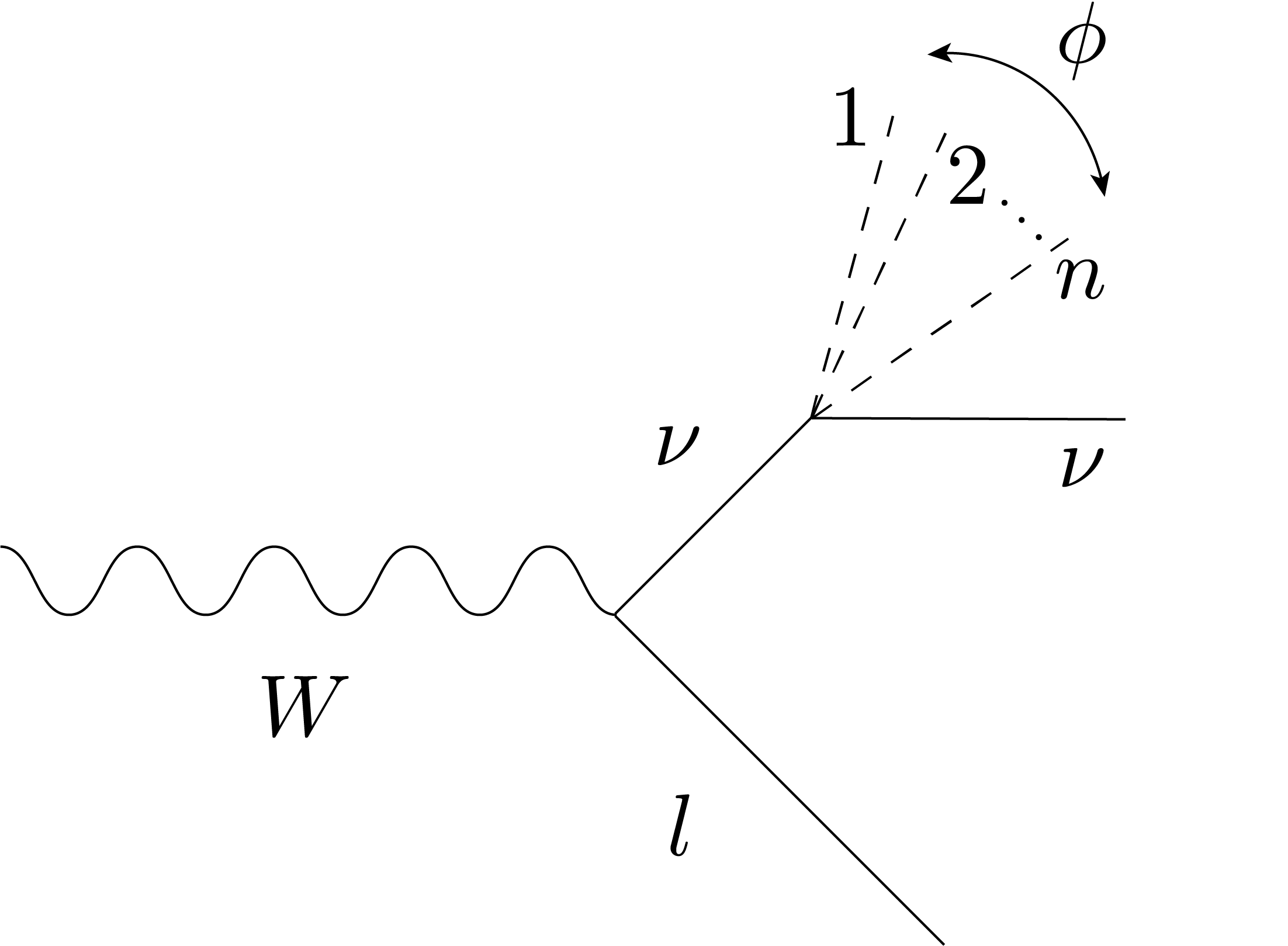}
    \caption{The decay $W\rightarrow n\phi+l+\nu$.}
    \label{fig:w_decay}
\end{figure}

The relevant process is shown in Fig.~\ref{fig:w_decay}.
We denote the four momenta of the $W$ boson, the outgoing lepton $l$, the intermediate neutrino and the outgoing neutrino as $p_W$, $p_l$, $q$ and $p_\nu$, respectively. We also denote the collective momenta of the $n$ Goldstone bosons $\phi$ as $p_\phi=p_1+p_2+\cdots+p_n$. We will ignore the masses of $\phi,\nu$ and $l$ throughout our calculation, except those appearing in the neutrino-Goldstone boson coupling. We will consider a single neutrino flavor, and later generalize to the standard three-flavor structure. The matrix element is
\begin{align}
i\mathcal{M}=-\left(\frac{g_2}{2\sqrt{2}}\right) \bar{u}(p_l)\gamma_\alpha(1-\gamma_5)\epsilon^\alpha \frac{i\slashed{q}}{q^2} \kappa_n v(p_\nu),  
\end{align}
where $\kappa_n$ denotes the coupling of neutrinos to $n$-Goldstone bosons, obtained from Eq.~\eqref{eq: interactions}. We see that $\kappa_n\propto\gamma_5$ for odd values of $n$. However, it can be checked that $i\mathcal{M}$ is the same for both even and odd values of $n$.
The amplitude can be squared to give
\begin{align}
\frac{1}{3}\sum|\mathcal{M}|^2= \frac{2}{3}\left(\frac{g_2\kappa_n}{2\sqrt{2}}\right)^2
\text{Tr}\left[\gamma^\beta \slashed{p}_l\gamma^\alpha\slashed{q} \slashed{p}_\nu \slashed{q}(1+\gamma_5)\right]\frac{1}{q^4}\Pi_{\alpha\beta},
\end{align}
where $\Pi_{\alpha\beta}=\left(-g_{\alpha\beta}+\frac{p_{W\alpha}p_{W\beta}}{M_W^2}\right)$. The factor of $1/3$ comes from averaging over initial $W$ polarizations. After some algebra this can be reduced to,
\begin{align}
\frac{1}{3}\sum|\mathcal{M}|^2=\left(\frac{g_2\kappa_n}{2\sqrt{2}}\right)^2\frac{1}{3q^4}
\left(8g^{\alpha\beta}\left[q^2(p_l\cdot p_\nu)-2(q\cdot p_\nu)(q\cdot p_l)\right]+32(q\cdot p_\nu)p_l^\alpha q^\beta-16q^2p_l^\alpha p_\nu^\beta\right)\Pi_{\alpha\beta}.
\end{align}
Let us now integrate over the phase space of the $n$ $\phi$ and the outgoing neutrino. To this end, we will use the following identities involving $k-$body phase space of massless particles~\cite{Abu-Ajamieh:2020yqi},
\begin{gather}
\Phi_k(P)=\int d\Phi_k(P) \equiv \int\frac{d^3p_1}{(2\pi)^3}\frac{1}{2E_1}\cdots \frac{d^3p_k}{(2\pi)^3}\frac{1}{2E_k} (2\pi)^4\delta^4(p_1+\cdots+p_k-P)=\frac{1}{8\pi(k-1)!(k-2)!}\left(\frac{E}{4\pi}\right)^{2k-4},\\
\int d\Phi_k(P)p_1^\mu=\frac{1}{k}\Phi_k(P)P^\mu,\\
\int d\Phi_k(P)p_1\cdot p_2=\frac{1}{2{k\choose2}}\Phi_k(P)P^2,
\end{gather}
with $E=\sqrt{P^2}$ being the center of mass energy. Utilizing the above identities and doing the contraction with $\Pi_{\alpha\beta}$ we get,
\begin{align}
\int d\Phi_{n+1}(q)\left(\frac{1}{3}\sum|\mathcal{M}|^2\right)&=\frac{8}{3q^2}\left(\frac{g_2\kappa_n}{2\sqrt{2}}\right)^2 \left[(p_l\cdot q)+\frac{2(p_W\cdot q)(p_l\cdot p_W)}{M_W^2}\right]\frac{\Phi_{n+1}(q)}{(n+1)},\nonumber\\
&=\left(\frac{g_2\kappa_n}{2\sqrt{2}}\right)^2\frac{q^{2n-4}}{3\pi(n+1)!(n-1)!}\frac{1}{(4\pi)^{2n-2}}\left(3(p_W\cdot p_l)-\frac{2(p_W\cdot p_l)^2}{M_W^2}\right).
\end{align}
As a final step, we go to the rest frame of the $W$ and integrate over the lepton momenta to get,
\begin{align}
\Gamma(W\rightarrow l+\nu+n\phi)=\frac{g_2^2M_W^{2n-1}\kappa_n^2}{(4\pi)^{2n}}\frac{1}{16\pi(n!)^2(n+2)!(n-1)}\quad\text{for}~n>1.    
\end{align}
Here we have also multiplied by a factor of $1/n!$ to account for the $n$ identical $\phi$s in the final state.  The total decay width of the $W$-boson into a final state containing an arbitrary number of $\phi$s is then obtained after summing over $n$,
\begin{align}
\Gamma_{\text{BSM}}\equiv \sum_{n>1}\Gamma(W\rightarrow l+\nu+n\phi)=\frac{1}{16\pi\times 96}\frac{g_2^2m_\nu^2}{M_W}\left(\frac{M_Wg_X}{4\pi m_X}\right)^4 {}_2F_4\left(\{1,1\},\{2,3,3,5\},\left(\frac{M_Wg_X}{4\pi m_X}\right)^2\right).  
\end{align}
The $n=1$ case is special as there is a soft divergence leading to a log enhancement of the form $\log \left ( m_W/m_X \right )$.  As this log is only present for $n=1$, we conservatively neglect the $n=1$ contribution to the decay width.

\end{document}